\documentstyle[aps,prl,floats]{revtex}
\begin{document}
\draft

\twocolumn[\hsize\textwidth\columnwidth\hsize\csname @twocolumnfalse\endcsname

\title{Vierbein walls in condensed matter.}

\author{
 G.E. Volovik$^{1,2}$
~}
\address{
$^{1}$Low Temperature Laboratory, Helsinki University of
Technology\\
P.O.Box 2200, FIN-02015 HUT, Finland\\
$^{2}$L.D. Landau Institute for
Theoretical Physics, \\ Kosygin Str. 2, 117940 Moscow, Russia
}

\date{\today}
\maketitle

\begin{abstract}
The effective field, which plays the part of the vierbein in general
relativity, can
have topologically stable surfaces, vierbein domain walls, where the effective
contravariant metric is degenerate. We consider vierbein walls separating
domains
with the flat space-time which are not causally connected at the classical
level.
Possibility of the quantum mechanical connection between the domains is
discussed.

\end{abstract}

\

\pacs{PACS numbers:     }

\

] \narrowtext
%\twocolumn

\section{Introduction.}

In some classes of superfluids and superconductors there is an effective field
arising in the low-energy corner, which acts on quasiparticles as gravitational
field. Here we discuss topological solitons in superfluids and superconductors,
which  represent the vierbein domain wall. At such surface in the 3D space
(or at
the 3D hypersurface in 3+1 space)  the vierbein is degenerate, so that the
determinant of the contravariant metric $g^{\mu\nu}$ becomes zero on
the surface.  An example of the vierbein domain
wall has been discussed in Ref.\cite{JacobsonVolovik}
for the $^3$He-A film.   When such  vierbein  wall moves, it spilts into
a black hole/white hole pair, which experiences the quantum friction force due
to Hawking radiation \cite{JacobsonVolovik}. Here we discuss the stationary
wall,
which is topologically stable and thus does not experience any dissipation.
Such
domain walls, at which one of the three "speeds of light" crosses zero, can be
realized in other condensed matter too: in superfluid
$^3$He-B \cite{SalomaaVolovik}, in chiral $p$-wave superconductors
\cite{MatsumotoSigrist,SigristAgterberg}, and in $d$-wave superconductors
\cite{Volovik1997}.

In the literature two types of the walls were
considered: with degenerate $g^{\mu\nu}$ and with degenerate
$g_{\mu\nu}$\cite{Bengtsson}. The case of degenerate
$g_{\mu\nu}$ was discussed in details in \cite{Bengtsson,BengtssonJacobson}.
Both types of the walls could be generic. According to Horowitz
\cite{Horowitz}, for a dense set of coordinate transformations the generic
situation is the 3D  hypersurface where the covariant metric $g_{\mu\nu}$
has rank
3.

The physical origin of the walls with the degenerate metric
$g^{\mu\nu}$ in general relativity have been discussed by
Starobinsky
\cite{Starobinsky}. They can arise after inflation, if the inflaton field has a
$Z_2$ degenerate vacuum. The domain walls  separates the domains with  2
diferent
vacua of the inflaton field. The metric $g^{\mu\nu}$ can everywhere satisfy
the Einstein equations in vacuum, but at the considered surfaces the
metric $g^{\mu\nu}$ cannot be diagonalized as
$g^{\mu\nu}={\rm diag}(1,-1,-1,-1)$. Instead, on such surface the metric is
diagonalized as
$g^{\mu\nu}={\rm diag}(1,0,-1,-1)$ and thus cannot be inverted. Though
the space-time can be flat everywhere, the coordinate transformation cannot
remove
such a surface: it can only move the surface to infinity. Thus the system
of such
vierbein domain walls divides the space-time into domains which cannot
communicate with each other. Each domain is flat and infinite as viewed by
a local observer living in a given domain. In principle, the domains can have
different space-time topology, as is emphasized by Starobinsky
\cite{Starobinsky}.

Here we consider the vierbein walls separating the flat space-time domains,
which
classically cannot communicate with each other across the wall, and discuss the
quantum mechanical behavior of the fermions in the presence of the domain wall.

\section{Vierbein Domain Wall}

The simplest example of the vierbein walls we are interested in is provided
by the
domain wall in superfluid
$^3$He-A film which separates domains with opposite orientations of the
unit vector
$\hat {\bf l}$ of the orbital momentum of Cooper pairs: $\hat {\bf l} =\pm
\hat
{\bf z}$.  Here the $\hat{\bf z}$ is along the normal to the film.
The Bogoliubov-Nambu Hamiltonian for fermionic quasiparticles is
\begin{equation}
{\cal H}= \left({p_x^2 +p_y^2-  p_F^2\over 2m}\right)\tau^3 + {\bf
e}_1\cdot {\bf
p}
\tau^1 +{\bf e}_2\cdot {\bf
p}
\tau^2
\label{HGeneral}
\end{equation}
Here $\tau^a$ are $2\times 2$ matrices for the Bogoliubov-Nambu spin; ${\bf p}=
\hat{\bf x}p_x+ \hat{\bf y}p_y$ is the 2D momentum (for simplicity we
assume that the film is narrow so that the motion along the normal to the
film is quantized and only the motion along the film is free); the complex
vector
\begin{equation}
{\bf e}= {\bf e}_1 +i{\bf e}_2 ~,~\hat{\bf l}={{\bf e}_1\times{\bf e}_2\over
|{\bf e}_1\times{\bf e}_2|}=\pm \hat{\bf z}
\label{VectorOP}
\end{equation}
is the order parameter.
If one considers the first term in Eq.(\ref{HGeneral}) as the mass term, then
the vectors
${\bf e}_1$ and
${\bf e}_2$ play the part of the zweibein for the 2D motion in the film.
We assume the following order parameter texture in the wall:
\begin{equation}
{\bf e}_1(x)  =\hat{\bf x} c_x(x) ~,~{\bf e}_2 =\hat{\bf y} c_y(x)~.
\label{Texture}
\end{equation}
where the "speed of light" propagating along the axis $y$ is constant, while
the "speed of light" propagating along the axis
$x$ changes sign across the wall:
\begin{equation}
 c_y(x)=c_0~,~c_x(x)=   c_0\tanh {x\over d}  ~~,
\label{SpeedsLight}
\end{equation}
 At $x=0$ the zweibein is degenerate: the vector product ${\bf e}_1\times{\bf
e}_2=0$ so that the $\hat {\bf l}$ vector is not determined.

Since the momentum projection $p_y$ is the conserved quantity, we come to
pure 1+1
motion. Further we assume that  (i) $p_y= \pm p_F$; and (ii) the parameters
of the
system are such that the thickness $d$ of the domain wall is large: $d
\gg
\hbar\over mc_0$. This allows us to consider the range of the momentum
$\hbar/d \ll p_x \ll mc_0$, where the term
$p_x^2$ can be either neglected as compared to the linear term or considered in
the semiclassical approximation.
Then rotating the Bogoliubov
spin and neglecting the noncommutativity of the $p_x^2$ term and $c(x)$ one
has the
following Hamiltonian for the 1+1 particle:
\begin{eqnarray}
{\cal H}= M( {\cal P})\tau^3  +  {1\over 2} (c(x) {\cal P}  + {\cal P}
c(x)) \tau^1~,\\    M^2( {\cal P}) ={{\cal P}^4 \over
4m^2}  + c_0^2p_y^2~.
\label{DiracCorrected}
\end{eqnarray}
where the momentum operator ${\cal P}_x
=-i\partial_x$ is introduced.
If the ${\cal P}^2$ term is completely neglected, one obtains the 1+1 Dirac
fermions
\begin{eqnarray}
{\cal H}=M\tau^3  +  {1\over 2} (c(x) {\cal P}  + {\cal P}
c(x)) \tau^1~,\\   M^2 = M^2( {\cal P}=0)= c_0^2p_y^2~.
\label{Dirac}
\end{eqnarray}
The
classical spectrum of quasiparticles,
\begin{equation}
E^2 - c^2(x)p_x^2=M^2 ~,
\label{E}
\end{equation}
corresponds to the  contravariant metric
\begin{equation}
g^{00}=1~,~g^{xx}=- c^2(x)~~.
\label{metric}
\end{equation}
The line element of the effective space-time is
\begin{equation}
ds^2= dt^2   - \bigl(c(x)\bigr)^{-2}\, dx^2  ~.
\label{LineElement}
\end{equation}
The metric element $g_{xx}$ is infinite at $x=0$.

The Eq.(\ref{LineElement}) represents a {\it flat} effective spacetime for any
function
$c(x)$. However the singularity at $x=0$, where $g_{xx}=\infty$, cannot be
removed
by the coordinate transformation. If at $x>0$ one introduces a
new coordinate
$\xi=\int dx/c(x)$, then the line element takes the standard  flat form
\begin{equation}
ds^2=dt^2 - d\xi^2  ~.
\label{LineElementFlat}
\end{equation}
However, the other domain -- the half-space with $x<0$ -- is completely
removed by
such transformation.  The situation is thus the
same as discussed by Starobinsky for the domain wall in the inflaton field
\cite{Starobinsky}.

The two flat spacetimes are disconnected in the relativisttic
approximation. However this approximation breaks down
near $x=0$, where the "Planck energy physics" becomes important and
nonlinearity in the energy spectrum appears in Eq.(\ref{DiracCorrected}):
The two
halves actually communicate due to the high-energy quasiparticles, which are
superluminal and thus can propagate through the wall.

\section{Fermions across Vierbein Wall.}

In classical limit the low-energy relativistic quasiparticles do not
communicate across the vierbein wall, because the speed of light $c(x)$
vanishes at $x=0$. However the quantum mechanical connection can be
possible.
There are two ways to treat the problem. In one approach one makes the
coordinate transformation first. Then in one of the domains, say, at $x>0$,
the line element is Eq.(\ref{LineElementFlat}), and one comes to the standard
solution for the Dirac particle propagating in flat space:
\begin{eqnarray}
\nonumber\chi(\xi)=
{A\over \sqrt{2}}\exp\left( i\xi\tilde E
\right)\left(\matrix{Q
\cr
Q^{-1}  \cr}\right) +\\
{B\over \sqrt{2}}\exp\left(-
i\xi\tilde E \right)\left(\matrix{Q
\cr
-Q^{-1} \cr} \right) ~,
\label{DiracSolutionFlat}\\
\tilde E =\sqrt{E^2-M^2}~,~Q=\left({E+M\over E-M}\right)^{1/4}~.
\end{eqnarray}
Here $A$ and $B$ are arbitrary constants.
In this approach it makes no sense to discuss any connection to the other
domain, which simply does not exist in this representation.

In the second approach we do not make the coordinate transformation and work
with both domains. The wave function for the Hamiltonian Eq.(\ref{Dirac}) at
$x>0$ follows from the solution in Eq.(\ref{DiracSolutionFlat}) after restoring
the old coordinates:
\begin{eqnarray}
\nonumber\chi(x>0)=\\
\nonumber
{A\over \sqrt{2c(x)}}\exp\left( i\xi(x)\tilde E
\right)\left(\matrix{Q
\cr
Q^{-1}  \cr}\right) +\\
{B\over \sqrt{2c(x)}}\exp\left(-
i\xi(x)\tilde E \right)\left(\matrix{Q
\cr
-Q^{-1}  \cr} \right),\\
\xi(x)=\int^x {dx\over
c(x)}
\label{DiracSolution+}
\end{eqnarray}
The similar solution exists at $x<0$.
We can now connect the solutions for the right and left half-spaces using
(i) the
analytic cotinuation across the point $x=0$; and (ii) the conservation of the
quasiparticle current across the interface. The quasiparticle current e.g.
at $x>0$
is
\begin{equation}
j =  c(x)  \chi^\dagger \tau^1 \chi=   |A|^2
-|B|^2  ~.
\label{Current}
\end{equation}

The analytic cotinuation depends on the choice of the contour  around the
point $x=0$ in the complex $x$ plane. Thus starting from
Eq.(\ref{DiracSolution+})
we obtain two possible solutions at $x<0$. The first solution is obtained
when the
point $x=0$ is shifted to the lower half-plane:
\begin{eqnarray}
\nonumber\chi^{I}(x<0)=\\
\nonumber
{-iA e^{ -{\tilde E\over 2T_H}}\over \sqrt{2|c(x)|}}\exp\left( i\xi(x)\tilde E
\right)\left(\matrix{Q
\cr
Q^{-1}  \cr}\right) +\\
{-iB e^{ { \tilde E\over 2T_H}}\over \sqrt{2|c(x)[}}\exp\left(-
i\xi(x)\tilde E \right)\left(\matrix{Q
\cr
-Q^{-1}  \cr} \right),
\label{DiracSolution-I}
\end{eqnarray}
where   $T_H$ is
\begin{equation}
T_H= {\hbar\over 2\pi}~ {dc\over dx} \bigg|_{x=0} ~.
\label{HawkingT}
\end{equation}
The conservation of the quasiparticle current (\ref{Current})
across the point $x=0$ gives the connection between parameters $A$ and
$B$:
\begin{equation}
|A|^2
-|B|^2 =|B|^2e^{ {\tilde E\over T_H}} - |A|^2e^{ -{\tilde E\over T_H}}~.
\label{CurrentConsI}
\end{equation}

The quantity $T_H$ looks like the Hawking
radiation temperature determined at the singularity. As follows from
Ref.\cite{JacobsonVolovik} it is the limit of the Hawking temperature when the
white hole and black hole horizons in the moving wall merge to form the static
vierbein wall. Note, that there is no real radiation when the wall does not
move.
The parameter $T_H/\tilde E \sim \partial \lambda /\partial x
$, where $\lambda= 2\pi/p_x=(2\pi /\tilde E) \partial c /\partial x$ is the de
Broglie wavelength of the quasiparticle. Thus the quasiclassical
approximation holds
if $T_H/\tilde E \ll 1$.

The second solution is obtained when the point $x=0$ is shifted to the upper
half-plane:
\begin{eqnarray}
\nonumber\chi^{II}(x<0)=\\
\nonumber
{iA e^{ {\tilde E\over 2T_H}}\over \sqrt{2|c(x)|}}\exp\left( i\xi(x)\tilde E
\right)\left(\matrix{Q
\cr
Q^{-1}  \cr}\right) +\\
{iB e^{ -{\tilde E\over 2T_H}}\over \sqrt{2|c(x)[}}\exp\left(-
i\xi(x)\tilde E \right)\left(\matrix{Q
\cr
-Q^{-1}  \cr} \right),
\label{DiracSolution-II}
\end{eqnarray}
and the current conservation gives the following relation between
parameters $A$
and
$B$:
\begin{equation}
|A|^2
-|B|^2 =|B|^2e^{ -{\tilde E\over T_H}} - |A|^2e^{ {\tilde E\over T_H}}~.
\label{CurrentConsII}
\end{equation}

Two solutions, the wave functions  $\chi^{I}$ and $\chi^{II}$, are connected by
the relation
\begin{equation}
\chi^{II}\propto\tau_3(\chi^{I})^*
\label{parity}
\end{equation}
which follows from the symmetry
 of the Hamiltonian
\begin{equation}
H^*=\tau_3 H \tau_3
\label{conjugation}
\end{equation}
The general solution is the linear combination of $\chi^{I}$ and $\chi^{II}$

Though on the classical level the two worlds on both
sides of the singularity are well separated, there is a quantum mechanical
interaction between the worlds across the vierbein wall.  The wave functions
across the wall are connected by the relation $\chi(-x)=\pm i\tau_3\chi^*(x)$
inspite of no possibility to communicate in the relativistic regime.

\section{Nonlinear (nonrelativistic) correction.}

In the above derivation we relied upon the analytic continuation and on the
conservation of the quasiparticle current across the wall. Let us justify
this using
the nonlinear correction in Eq.(\ref{DiracCorrected}), which was neglected
before.
We shall work in the quasiclassical approximation, which holds if $\tilde E\gg
T_H$. In a purely classical limit one has the dispersion
\begin{equation}
E^2= M^2 + c^2(x)p_x^2+{p_x^4\over 4m^2} ~,
\label{NonlineaClassical}
\end{equation}
which determines two classical trajectories
\begin{equation}
p_x(x)= \pm \sqrt{2m\left(\sqrt{\tilde E^2+m^2c^4(x)} -mc^2(x)\right)} ~.
\label{NonlineaClassicalTrajectories}
\end{equation}
It is clear that there is no singularity at $x=0$, the two trajectories
continuously
cross the domain wall in opposite directions, while the Bogoliubov spin
continuously
changes its direction. Far from the wall these two trajectories give the two
solutions,  $\chi^{I}$ and $\chi^{II}$, in the quasiclassical limit $\tilde
E\gg
T_H$.  The function $\chi^{I}$
\begin{eqnarray}
\chi^{I}(x>0)={1\over \sqrt{2|c(x)|}}\exp\left( i\xi(x)\tilde E
\right)\left(\matrix{Q
\cr
Q^{-1}  \cr}\right),  \\
\chi^{I}(x<0)={-i\over \sqrt{2|c(x)|}}\exp\left( -i\xi(x)\tilde
E
\right)\left(\matrix{Q
\cr
-Q^{-1}  \cr}\right).
\label{QuasiclassicalSolutionI}
\end{eqnarray}
describes the propagation of the quasiparticle from
the left to the right without reflection at the wall: in the quasiclassical
limit reflection is suppressed. The function $\chi^{II}$ desribes the
propagation
in the opposite direction:
\begin{eqnarray}
\chi^{II}(x>0)={1\over \sqrt{2|c(x)|}}\exp\left( -i\xi(x)\tilde E
\right)\left(\matrix{Q
\cr
-Q^{-1}  \cr}\right),  \\
\chi^{II}(x<0)={i\over \sqrt{2|c(x)|}}\exp\left( i\xi(x)\tilde
E
\right)\left(\matrix{Q
\cr
Q^{-1}  \cr}\right),
\label{QuasiclassicalSolutionII}
\end{eqnarray}
The quasiparticle
current far from the wall does obey the Eq.(\ref{Current}) and is conserved
across
the wall. This confirms the quantum mechanical connection between the spaces
obtained in previous section.

In the limit of small mass $M\rightarrow 0$, the particles become chiral
with the
spin directed along or opposite to the momentum $p_x$. The spin structure
of the
wave function in a semiclassical approximation is given by
\begin{equation}
\chi(x)= e^{i\tau_2{\alpha\over 2}}  \chi(+\infty) ~,~\tan \alpha= {p_x\over
2mc(x)}.
\label{SpinRotation}
\end{equation}
Since $\alpha$ changes by $\pi$ across the wall, the spin of
the chiral quasiparticle rotates by $\pi$: the righthanded particle
tranforms to the lefthanded one when the wall is crossed.

\section{Discussion.}

It appears that there is a quantum mechanic
coherence between the two flat worlds,  which do not interact classically
across
the vierbein wall. The coherence is established by  nonlinear  correction
to the spectrum of chiral particle: $E^2(p)=c^2p^2 + ap^4$.  The
parameter $a$ is positive in the condensed matter analogy, which  allows the
superluminal propagation across the wall at high momenta $p$. But the
result does
not depend on the magnitude of
$a$: in the relativistic low energy limit the amplitudes of the
wave function on the left and right sides of the wall remain equal in
the quasiclassical approximation, though in the low
energy corner the communication across the wall is classically forbidden.
Thus the only relevant input of
the "Planck energy" physics is the mere possibility of the superluminal
communication  between the worlds across the wall.   That is why the
coherence between  particles propagating in two classically disconnected
worlds can be obtained even in the relativistic domain, by using the analytic
continuation  and the conservation of the particle current across the
vierbein wall.

\acknowledgements

I thank A. Starobinsky for illuminating discussions. This work was supported in
part by the
Russian Foundations for Fundamental Research grant No. 96-02-16072
and by
European Science Foundation.

\end{document}